\documentclass[namedreferences]{solarphysics}
\usepackage[hyperref,optionalrh]{spr-sola-addons} 
\usepackage{graphicx}        
\usepackage{url}        
\usepackage[usenames,dvipsnames]{color} 
\usepackage[normalem]{ulem}     
\usepackage{pdflscape}
\usepackage{chemfig}

\newcommand{\fig}[1]{Figure~\ref{fig_#1}}

\newcommand{\BE}{\begin{equation}}
\newcommand{\EE}{\end{equation}}
\newcommand{\BA}{\begin{eqnarray}}
\newcommand{\EA}{\end{eqnarray}}

\newcommand{\arcsec}{''} 
\newcommand{\degree}{^\circ} 

\newcommand{\eg}{\textit{e.g.,}}

\chardef\us=`\_

\begin{document}

\begin{article}
\begin{opening}

\title{
{Sympathetic Quiet and Active Region Filament Eruptions} }
\author[addressref={aff1},corref,email={kkoleva@space.bas.bg}]{\inits{Kostadinka}\fnm{Kostadinka }~\lnm{Koleva}}

\author[addressref={aff2}, email={setiapooja.ps@gmail.com}]{\inits{Pooja}\fnm{Pooja}~\lnm{Devi}}
\author[addressref={aff2},email={rchandra.ntl@gmail.com}]{\inits{Ramesh}\fnm{Ramesh}~\lnm{Chandra}}
\author[addressref={aff2}, email={reetikajoshi.ntl@gmail.com}]{\inits{Reetika}\fnm{Reetika}~\lnm{Joshi}}
\author[addressref={aff3}, email={duchlev@astro.bas.bg}]{\inits{Peter}\fnm{Peter}~\lnm{Duchlev}}
\author[addressref={aff3}, email={mdechev@astro.bas.bg}]{\inits{Momchil}\fnm{Momchil}~\lnm{Dechev}}

\address[id=aff1]{ Space Research and Technology Institute,  Bulgarian Academy of Sciences, Bulgaria}
\address[id=aff2]{Department of Physics, DSB Campus, Kumaun University, Nainital 263 002, India}
\address[id=aff3]{Institute of Astronomy and NAO, Bulgarian Academy of Sciences, Bulgaria}

\runningauthor{K. Koleva et al.}
\runningtitle{Sympathetic Filament Eruptions}

\begin{abstract}
We present the  observations of three sympathetic filament eruptions occurring on 19 July 2015 namely F$_1$, F$_2$, and F$_3$. The events were observed in UV/EUV wavelengths by Atmospheric Imaging Assembly onboard the Solar Dynamics Observatory and by Global Oscillation Network Group telescope in H$\alpha$ line.
As filament F$_1$ starts to erupt, a part of it falls close to the location of the F$_2$ and F$_3$ filaments. This causes the eruption of F$_2$ and F$_3$ during which the two filaments merge together and trigger a medium-class CME and a long-duration GOES C2.1 class flare. We discuss the dynamics and kinematics of these three filament eruptions and related phenomena.

\end{abstract}

\end{opening}

\section{Introduction}
\label{sect_Introduction}

Solar filaments/prominences are well-known phenomenon in the solar atmosphere.
They present a variety of cool and dense objects, ranging from long-lived quiescent filament/prominence to short-lived active region filament/prominence.
Their nature is described in many studies
\citep{Ballegooijen1989, Chae2001, Labrosse2010, Mackay2010, Schmieder2013, Gibson2018}. It is believed that they are supported in magnetic dips \citep{Aulanier2002a, Mackay2010, Gibson2018}, observed on polarity inversion line (PIL).

When the balance between magnetic pressure and magnetic tension in the filaments becomes unstable by any kind of mechanism, they can erupt. 
Observations show that based on the relation between the filament mass and corresponding supporting magnetic structure
filaments can erupt fully \citep{Gopalswamy2003, Schrijver2008, Chandra2010} or partially \citep{Gibson2006, Joshi2014, Cheng2018, Monga2021} and sometimes the eruption can be 
failed \citep[\eg][]{Liu2009, Kumar2011, Joshi2013}. 
The full or partial eruption are usually associated with the coronal mass ejections (CMEs), which later on become Interplanetary CME (ICME), responsible for space weather disturbances \citep{Gopalswamy2007, Schmieder2020}.

Sometimes, merging of the filaments is also observed 
\citep{Schmieder2004, Chandra2011, Jiang2014, Luna2017}.
Merging can tell us about the formation and the dynamical evolution of the filaments.
Cases are reported, where two filaments 
merge and the result can be stable or eruptive filament. Such phenomena were simulated in high beta plasma condition by \inlinecite{Linton2006} and in low beta plasma coronal condition by \inlinecite{Aulanier2006}  and \inlinecite {Torok2011}. In Addition to it laboratory experiment has also been performed \citep{Gekelman2012}.

Occasionally, the eruptions occur in a short interval of time at same or different locations on the solar surface \citep{Biesecker2000, Moon2002, Wang2002, Zhukov2007, Liu2009}.
The consecutive eruptions, occurring in the same active region within a relatively short time interval, 
are defined as sympathetic eruptions.
Other cases could also be recognized as sympathetic eruptions, such as the recurring/successive eruptions that appear at  different locations of the solar surface. Such events can occur in both quiet and active regions \citep{Moon2002, Wheatland2006, Schrijver2011}. Sympathetic eruptions have already been observed in the past 
\citep{Richardson1936,Richardson1951, Becker1958}  and 
it is believed that they can be physically connected by the coronal loops.
It was found that in case of sympathetic eruptions, the multiple flux systems erupt. First, the eruption starts in one active region, pushing the overlaying magnetic flux and causing other flux systems to erupt \citep{Delannee1999, Wang2002, Liu2006,Zhukov2007}. Another possibility was suggested by \inlinecite{Khan2000}. According to them, the propagation of EUV waves can destabilize the adjacent loop system and ultimately lead to another eruption.

The sympathetic eruptions were modelled in MHD numerical simulations. \inlinecite{Ding2006} performed the 2.5 D time-dependent MHD model. In this model they scrutinized the catastrophic behavior of a multiple flux rope system, which contains three magnetic flux ropes  in three sets of  separate loop arcades. They concluded that the eruption of the first flux rope disturbed the stability of the second and third flux ropes and forced them to erupt.
According to the model proposed by \inlinecite{Wheatland2006},
if a flare occurs in the location of a separator, it temporary increases the probability of flaring at all separators (a complex of reconnecting structures). Recently \inlinecite{Torok2011} presented 3 D MHD simulation of two magnetic flux ropes and reproduced the 2010 August 1 quiet filament eruptions. Their results support the hypothesis that the trigger mechanisms of sympathetic eruptions can be related to the  large-scale coronal magnetic field. 
Despite the numerous observations and simulations, the exact reason for the sympathetic eruptions is still not well understood.

In this paper we present the observations of three sympathetically erupting filaments. Each eruption was associated with a CME. In addition to this, the observations also show the merging of two filaments. The paper is organized as follows: Section \ref{sect_Observations}
describes the data sets used in the study.  The results of our analysis are presented in Section \ref{sect_results}. Finally, the discussion and summary of the investigation are presented in Section \ref{discussion}.

\section{Observational Data Sets}
\label{sect_Observations}
For this study, we used the data from following sources:

\begin{enumerate}
\item {\bf SDO/AIA data:} For the evolution and dynamics/kinematics study of the filament eruptions, we used data from the Atmospheric Imaging Assembly (AIA; \opencite{2012SoPh..275...17L}) onboard the Solar Dynamics Observatory (SDO; \inlinecite{Pesnell2012}). The AIA consists of seven Extreme Ultra-Violet (EUV) and three Ultra-Violet (UV) channels which probe the solar corona with a pixel resolution of 0.6$\arcsec$ and an average cadence of 12~s. The AIA image field-of-view (FOV) reaches 1.3 solar radii. For the present study we used 1 min cadence data from the AIA 171, 193 and 304 \AA\ channels. \\

\item {\bf H$\alpha$ Data: } 
H$\alpha$ images from Global Oscillation Network Group (GONG; \opencite {1996Sci...272.1284H}) were used to study the chromospheric evolution of filament eruptions. GONG observes the full Sun in H$\alpha$ with a cadence of 1 min and a pixel resolution of 1$\arcsec$.

\medskip

\item {\bf SDO/HMI magnetic field data :}
The line-of-sight magnetograms taken by Helioseismic and Magnetic Imager (HMI; \opencite{2012SoPh..275..207S,Schou2012}) on board SDO were used to explore the photospheric magnetic fields configuration in the corresponding regions. The HMI LOS magnetograms used in this study have a cadence of 10~min and pixel size of 0.5$\arcsec$. The 1$\sigma$ noise level for HMI line-of-sight magnetogram is 10 G (\opencite {2012SoPh..279..295L}).
The HMI magnetograms and AIA images were co-aligned by using the UV AIA 1600 \AA\ images, which was consequently aligned with the AIA EUV channels. All data were corrected for projection effect and derotated to 23:20~UT on 18 July 2015.

\medskip
\item {\bf LASCO CME data:} 
The CME association of the erupted filaments was traced in the field-of-view (FOV) of the C2 Coronagraph (2.2 -- 6 R$_\odot$) of Large Angle and Spectrometric Coronagraph (LASCO; \opencite{Brueckner1895}) on board the SOHO satellite.

\end{enumerate}

\begin{figure*} 
\centering
\vspace{-1.0cm}
\includegraphics[width=\textwidth]{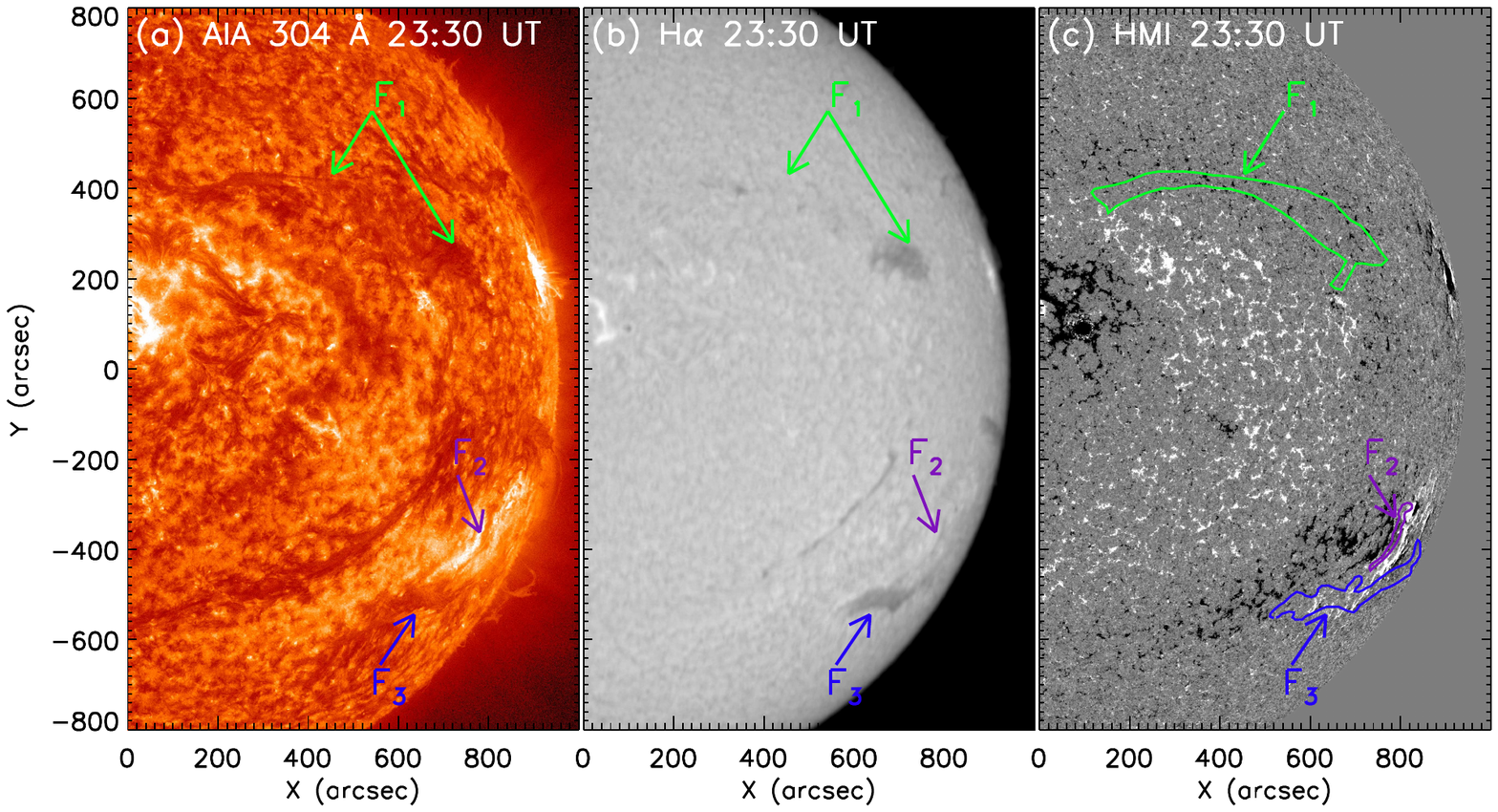}
\caption{Three visible filaments in AIA 304 \AA\ (panel a) and H$\alpha$ (panel b) on 18 July 2015. The contours of 304 \AA\ filaments F$_1$, F$_2$, and F$_3$ are over-plotted over HMI magnetogram (panel c) with green, purple, and blue colors, respectively.}
\vspace{-7.0cm}
\label{fig_304_hal_hmi}
\end{figure*}
\section{Results}
\label{sect_results}

On 19 July 2015, three filaments erupted sympathetically. First  one was a quiescent filament, while the other two filaments were situated in an active region. 
The dynamics and the kinematics of these eruptions are presented in the following 
sub-sections.

\begin{figure*}[t] 
\centering
\includegraphics[width=\textwidth]{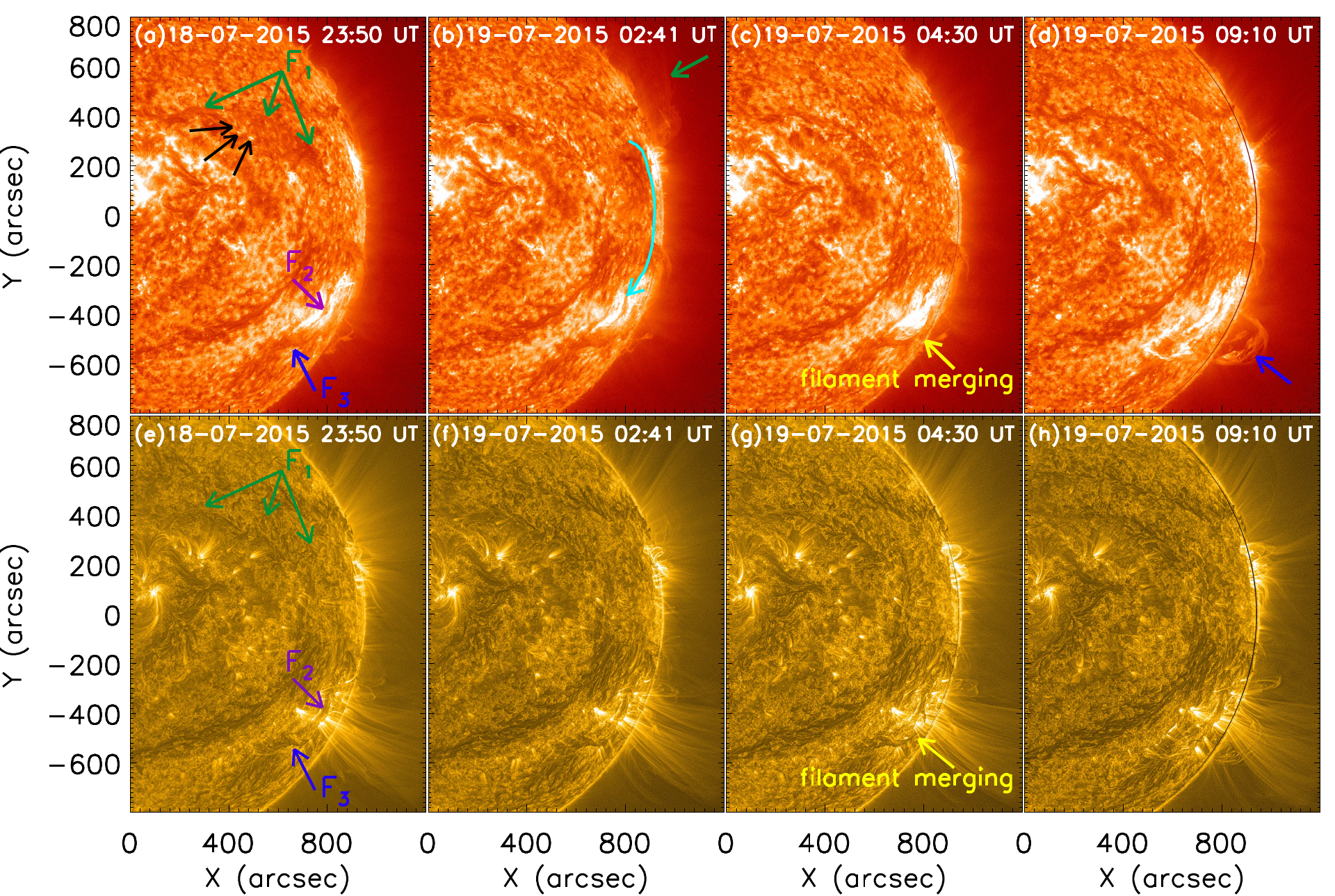}
\caption{Evolution of the event in AIA 304 \AA\ and 171 \AA\ wavelength in top and bottom rows, respectively. Panels (a) and (e) show the filaments F$_1$, F$_2$, and F$_3$ with green, purple, and blue arrows, respectively. Panel (b) shows the erupting filament F$_1$ with green arrow and cyan arrow shows the filament material falling downward (towards filament F$_2$), which triggers the filament F$_2$ (panel d). The yellow arrow in panels (c) and (g) is pointing the merging of filaments F$_2$ and F$_3$. The AIA 171 \AA\ images are processed with MGN method. The black arrows in panel (a) shows the pre-eruptive brightening in the vicinity of filament F$_1$. The Movies of these data are available in the Electronic Supplementary Materials.}
\label{fig_evolution304_171}
\end{figure*}

\subsection{Dynamics}

 We named the three filaments as F$_1$, F$_2$, and F$_3$ respectively. Filament F$_1$ was located in northern hemisphere, on the other hand filaments F$_2$, and F$_3$ were located in southern hemisphere.
The location of all these three filaments are shown in panels `a' and `b' of Figure~\ref{fig_304_hal_hmi} in AIA 304 \AA\ and H$\alpha$ respectively. In panel `c' of the figure the AIA 304 \AA\ filament contours are overlaid on photospheric magnetic field by green, pink and blue colours. This image  was obtained before the eruption on 18 July 2015. 
Filament F$_1$ was a large filament (projection length $\sim$ 450 Mm) that was observed on the solar disk  between 11 -- 19 July 2015. It survived around eight days on solar disk and  erupted on 19 July 2015.
The filaments F$_2$ and F$_3$ were located in the NOAA active region (AR) 12384. Initially these filaments were observed as a single filament up to 16 July 2015, which on 17 July 2015 splitted into two smaller parts as earlier named as F$_2$ and F$_3$.

Figure \ref{fig_evolution304_171} illustrates the filament eruption evolution in AIA 304 \AA\ and 171 \AA\ wavelengths in top and bottom rows, respectively.
To make the evolution more clear, we have created the MGN processed images in 171 \AA. This method is proposed by \citet{Morgan2014}.  It is based on the localized normalization of the data at different spatial levels. There are several parameters in this code, which can be changed according to the waveband {\it namely} `$\gamma$’ , `k', `h'.  The `$\gamma$’ parameter is useful for the global gamma transformation of the image, the parameter `k’ controls the sharpness of the gamma transformation and `h’ is the approximate weight of the global normalized image. We used the original default values of `$\gamma$’ and `k’ described in the original code as 3.2 and 0.7, respectively. We modified slightly the value of  `h’ to 0.9, as it can be changed for the type of input image and for the desired output. Before applying the MGN technique, firstly the AIA data is pre-processed using $aia\_prep$ procedure and all the images are aligned at a fixed time, to compensate the solar rotation effect, using $drot\_map$ routine available in SSWIDL. Such image processing 
is useful to present clearly the structural evolution of the eruption \citep{Devi2021}. 
Here we discuss the eruption evolution of the filaments step-by-step. The filament F$_1$ started to rise $\sim$~ 01:00 UT on 19 July 2015. The erupted material 
went into two major directions. Part of the erupted filament moved towards northwest direction. This erupted part 
became visible in LASCO C2 FOV  as a CME at $\sim$ 03:36 UT with a speed of $\sim$ 126 km s$^{-1}$.
As the filament erupted
two parallel elongated brightening along the PIL, where the F$_1$ filament was situated before its eruptions, were observed. These two ribbons were very faint. 
We could not see any enhancement in the GOES X-ray flux at this time. This could be due to the weak reconnection occurrence during the filament eruption as illustrated in the study of \inlinecite{Chandra2021}.

Remaining part of the erupted F$_1$ filament advanced into south direction, as shown by cyan arrows in \fig{evolution304_171}. Finally it fell down towards the filament F$_2$ and reached up to its north feet (see \fig{evolution304_171}(b)). 
As a result 
filament F$_2$ started to rise and merged with filament  F$_3$ around 04:33~UT. Part of the filament F$_2$ was skipped from the solar surface and appeared as a very faint CME at 05:24~UT in LASCO C2 FOV.

The merged filament started to rise slowly in southwest direction at $\sim$ 05:10~UT. This eruption was associated with a C-class GOES solar flare and a partial halo (width $\sim$ 194$\degree$) CME with a linear speed of $\sim$  782 km s$^{-1}$, which was first observed in the LASCO C2 FOV at $\sim$ 09:48~UT at a height of $\sim$ 2.9 R$_{sun}$.

The  chronology of these eruptions and related activities are presented in Table~1.

\begin{center}
\tabcolsep=0.07cm
\begin{table}[hbt!]
	\renewcommand{\arraystretch}{1.5}
\tiny
\begin{tabular*} {\textwidth} {ccc}
\hline
\centering
Time (UT) & Activities & Notes \\
\hline
23:40$^{*}$ -- 00:50 & Activation of filament F$_1$ & * time one day earlier\\
00:50 -- 03:20 & Part of  F$_1$ material moved northwest & -- \\
03:36 & CME appearance in LASCO FOV & Related with northwest part \\
01:30 -- 04:00 & Part of F$_1$ material moved southward  & Reached upto F$_2$ \\

04:11 & Activity in filament F$_2$ & Partial eruption \\
04:33 & Start of the merging of F$_2$ and F$_3$ filaments & -- \\
04:33 -- 05:10 & Small oscillations in F$_3$ & --\\
05:24        & small CME in LASCO FOV & Related to F$_2$ partial eruption \\
05:00 & Final merging of F$_2$ and F$_3$ filaments\\
05:10 & F$_3$ starts to rise
& Small velocity (few km s$^{-1}$ )\\
07:00 & F$_3$ starts faster velocity & velocity $\sim$ 12 km s$^{-1}$  \\
09:00 & F$_3$ acceleration phase &  velocity $\sim$ 138 km s$^{-1}$ \\
09:10 & GOES C2 flare onset & Long duration flare ($\sim$ 7 hrs) \\
09:48 & CME appearance in LASCO FOV & A partial halo CME\\
\hline
\end{tabular*}
\caption{Chronology of the eruptions. }
\label{tab_velocity}
\end{table}
\end{center}

\begin{figure*}        
\centering
\vspace{-0.1cm}
\includegraphics[width=\textwidth]{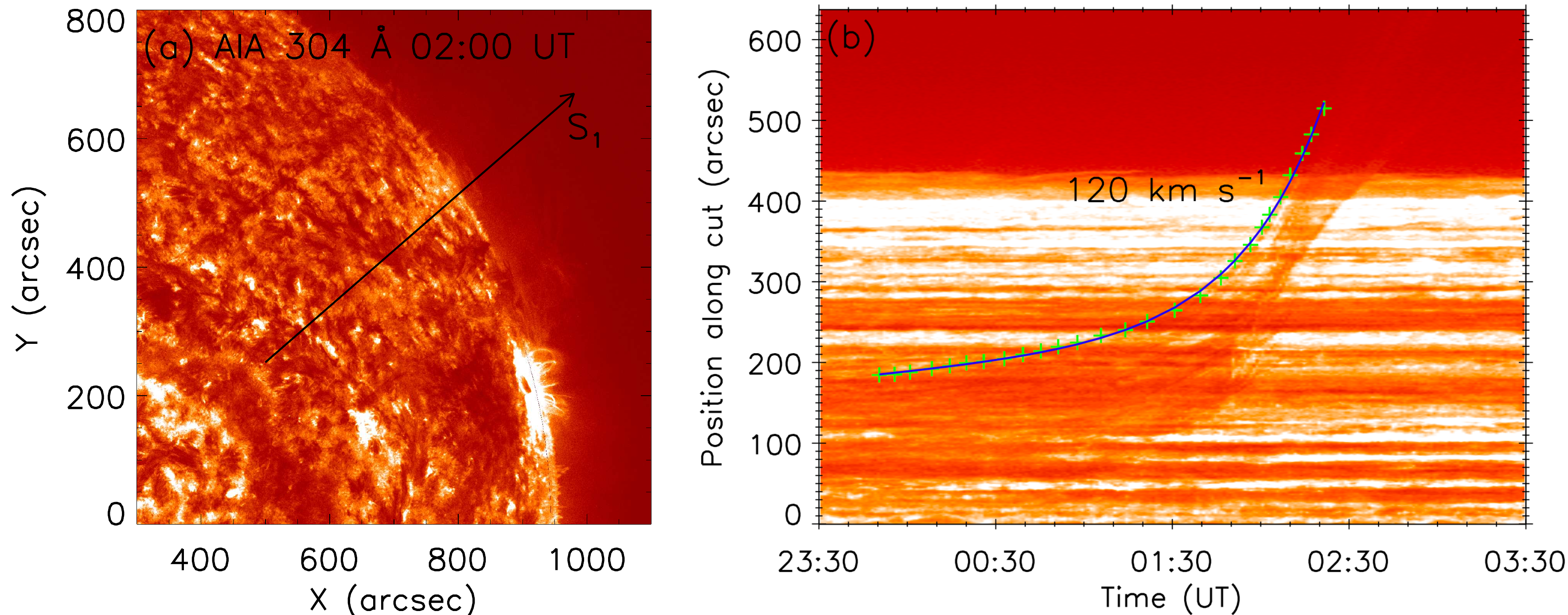}
\caption{Panel (a) is the AIA 304 \AA\ image from  19 July 2015 02:00 UT showing the direction of the slice S$_1$. Panel (b) shows the time-distance plot corresponding to the slice S$_1$. The green `plus' symbols in panel (b) are the data points chosen from the time-distance plot and blue solid line is the fitting curve to these data points. The fitted function is $a e^{b(t-t_0)}+ct+h_0$. The eruption speed  estimated from the fitting is found to be $\sim$ 120 km s$^{-1}$.}
\label{fig_slit_S1}
\end{figure*}

\subsection{Kinematics}

To investigate the temporal and spatial connection between the three erupted filaments, we performed the
time-distance analysis. This technique is based on the exploration of the motion of plasma material along an artificial slit. For this purpose, we have selected different slits in some selected directions. These slits were named as S$_1$, S$_2$, S$_3$, and S$_4$ respectively.

The slit S$_1$ was selected in order to analyse the kinematic behavior of
$F_1$ eruption in the north west direction. The selected slit and the corresponding time-distance plot are shown in Figure~\ref{fig_slit_S1}.  We have selected some points
in the edge of the time-distance slice and overplotted them in the same image (Figure~\ref{fig_slit_S1}b) with `plus' symbol in green color. Further, these data points were fitted by a combination of linear and exponential functions, namely ($a e^{b(t-t_0)}+ct+h_0$), as done by \citet{Cheng2020}.
Where a, b, c, and $h_0$ are arbitrary constants and the time $t_0$ is fixed at 23:50 UT on 18 July 2015. The fitted function is plotted as blue solid line in panel `b' of the Figure.
The computed speed is $\sim$ 120 $\pm$ 6 km s$^{-1}$. To determine the exact start time of the eruption, we used the equation $t_{start} = \frac{1}{b}$ln$(\frac{c}{ab}$). According to our results the eruption started at $\sim$ 00:42~UT on 19 July 2015.

\begin{figure*}      
\centering
\vspace{-2cm}
\includegraphics[width=\textwidth]{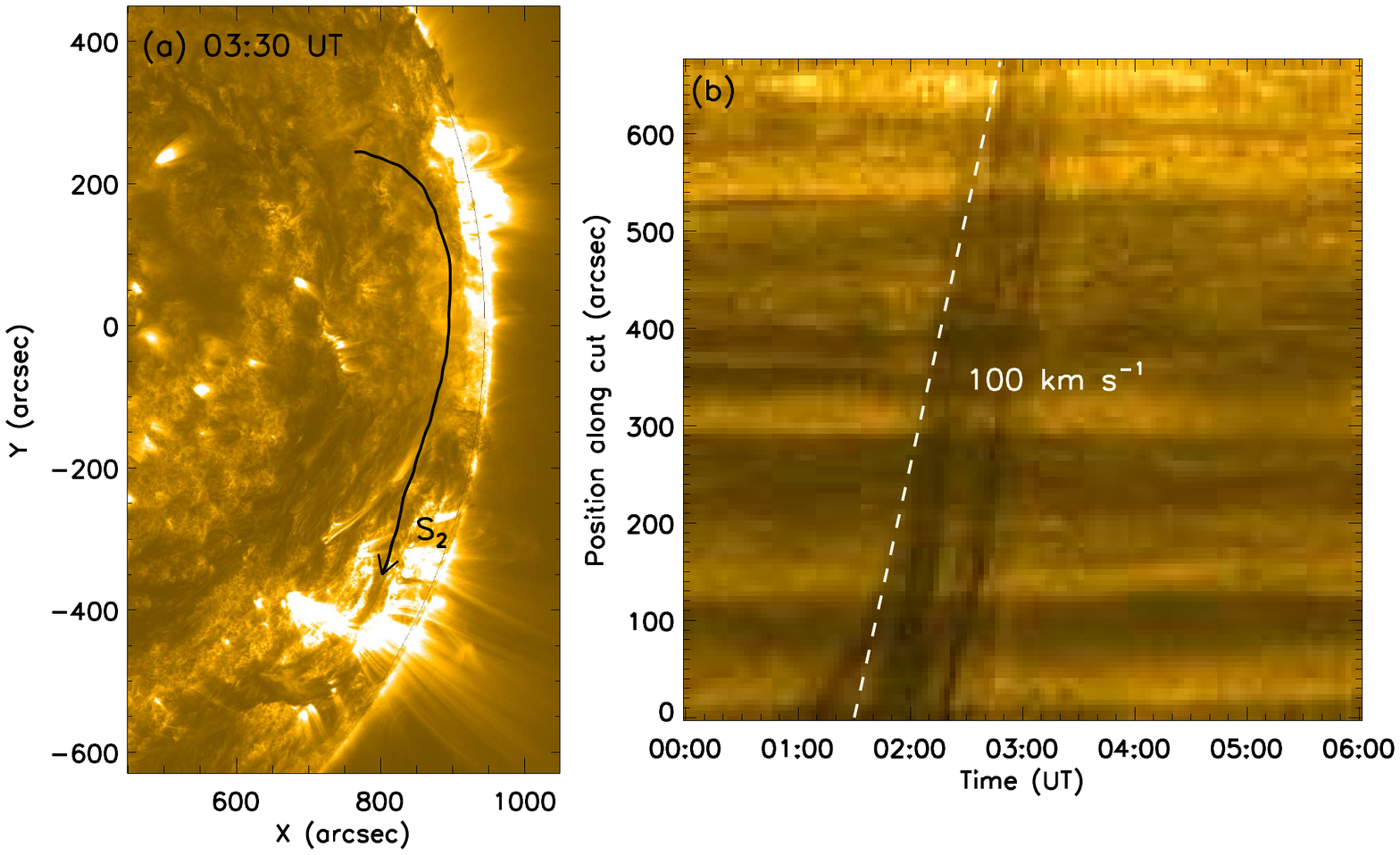}
\caption{Panel (a): Image of AIA 171 \AA\ on 19 July 2015 03:30 UT with the curved slice shown by black arrow, in the direction of the material falling from filament F$_1$ towards F$_2$. Panel (b): time-distance plot corresponding to the slice in panel (a).}
\vspace{-7.0cm}
\label{fig_slit_S2}
\end{figure*}

The slit S$_2$ was chosen at southward direction where part of F$_1$ material was observed to move.
The slit position and the time-slice is depicted in Figure~\ref{fig_slit_S2}. 
The speed of material going in this direction was computed using the straight-line fit. The estimated speed value is about 100 $\pm$ 2 km s$^{-1}$, which is comparable but slightly slower than the speed of filament in north-west direction. This slower speed could be due to the following possibilities: Due to long curved path of S$_2$ along the closed loop channel (evidenced by the PFSS extrapolation in Figure \ref{fig_pfss}), the material ejected from the filament F$_1$ decelerated and result as a slower speed. Another reason for the slower speed could be the expansion of the filament F$_1$. The erupting filament material reached the feet of filament F$_2$ at $\sim$ 03:20~UT. 

\begin{figure*}                          
\centering
\vspace{-0.1cm}
\includegraphics[width=\textwidth]{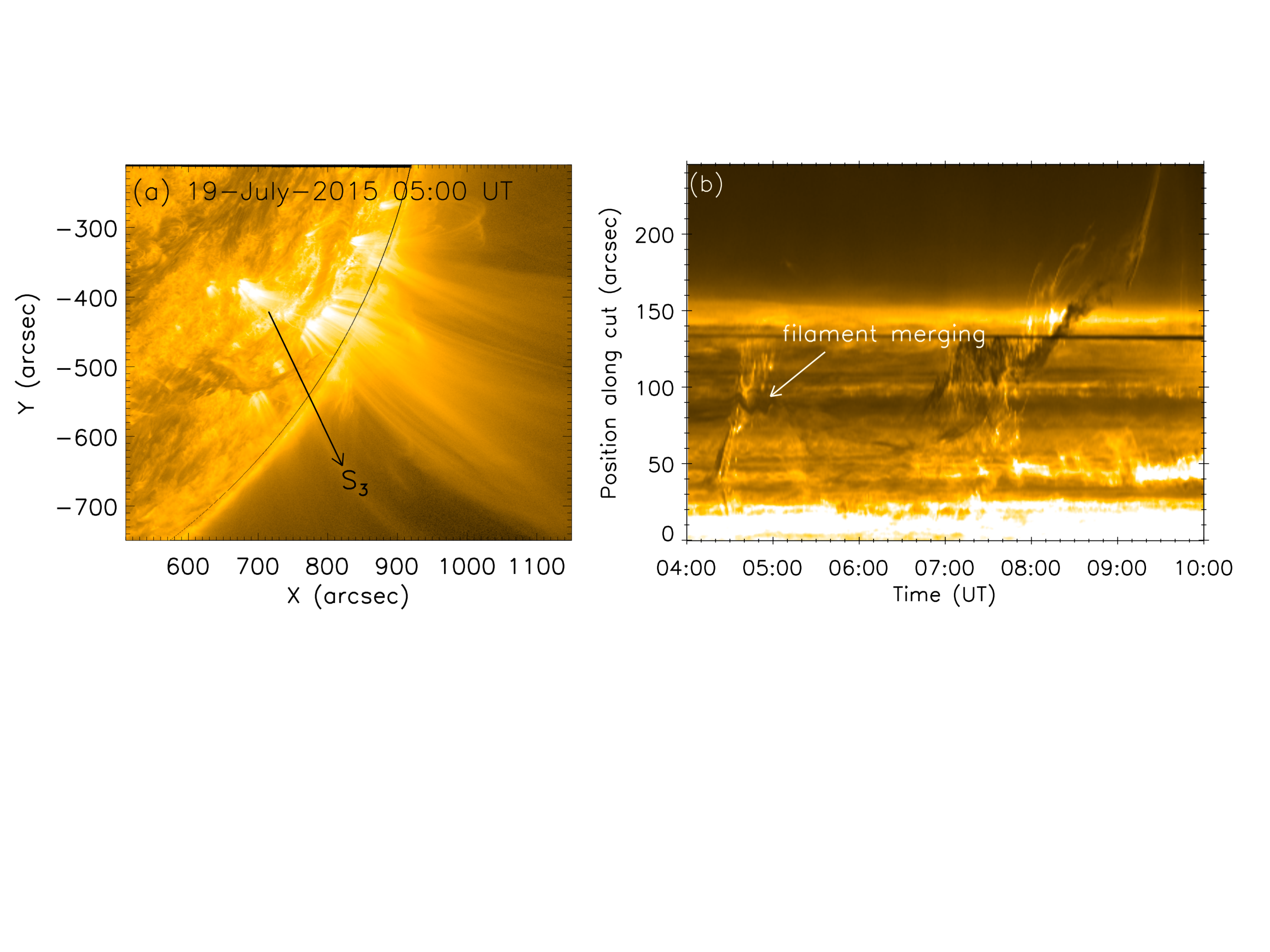}
\caption{Image of AIA 171 \AA\ showing the slice S$_3$ (panel a) and the time-distance plot corresponding to this slice in panel (b). The time-distance plot shows the merging of the filaments.}
\vspace{-3.0cm}
\label{fig_slit_S3}
\end{figure*}

The slit S$_3$ is placed between the filaments F$_2$ and F$_3$. The purpose of this slit is to comprehend the observed merging of these two filaments. The results are plotted in Figure~\ref{fig_slit_S3}. The time-distance plot indicates that the filament F$_2$ was in stationary state up to $\sim$ 04:10~UT and it started to rise at $\sim$ 04:20~UT. Around this time the filament F$_2$ became unstable due to the continuous flow of plasma material of F$_1$ filament. As a result the filament F$_2$  merged with F$_3$ at around 05:00~UT. 

\begin{figure*}                
\centering
\vspace{-0.1cm}
\includegraphics[width=\textwidth]{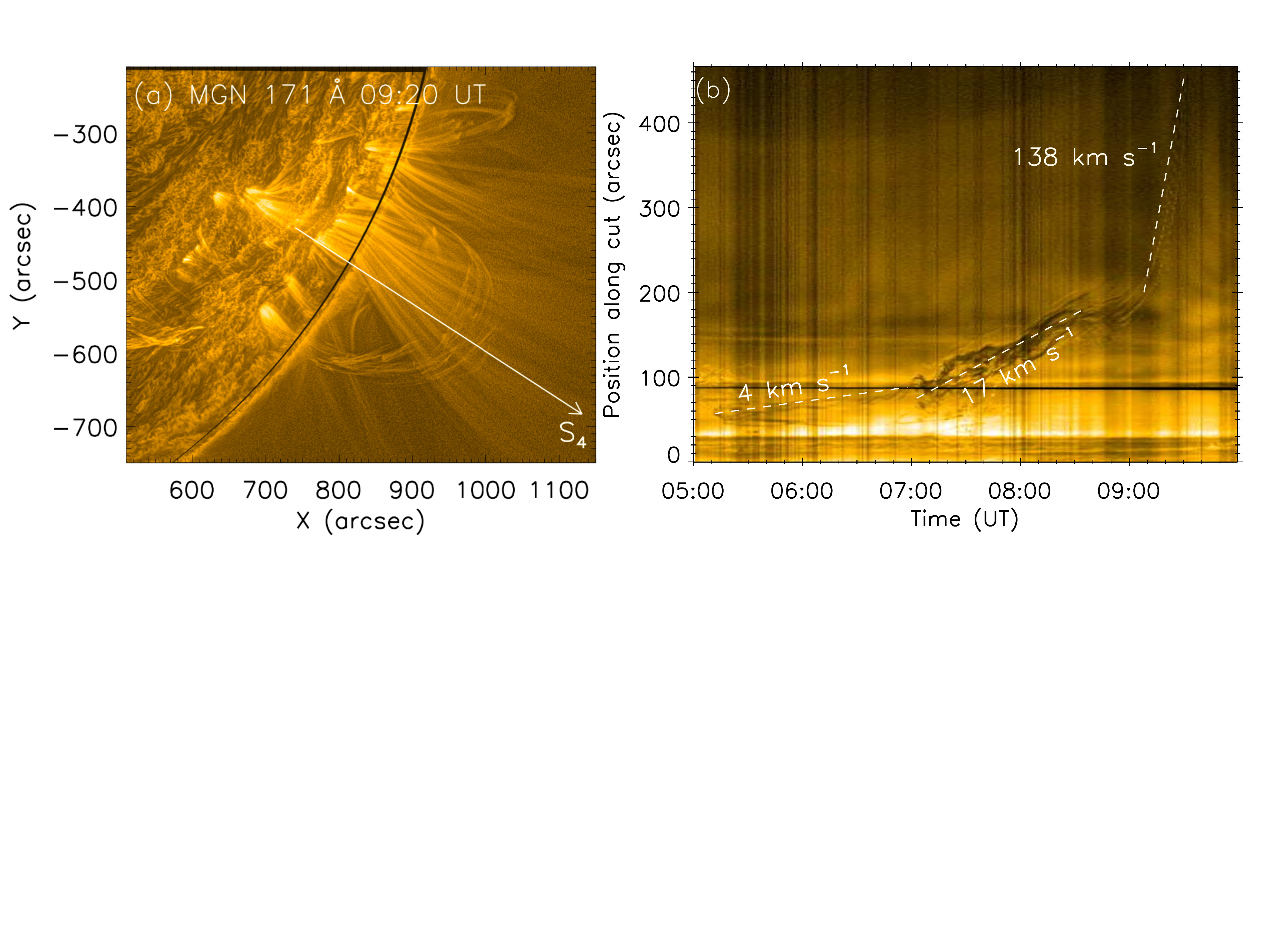}
\caption{Panel (a): AIA 171 \AA\ image, processed with the MGN, showing the  slice S$_4$ which was taken in the direction of F$_3$ eruption. 
Panel (b): The time-distance plot of F$_3$ eruption, showing that it erupted in three phases with speeds of 4 km s$^{-1}$, 17 km s$^{-1}$ and 138 km s$^{-1}$, respectively.}
\vspace{-3.0cm}
\label{fig_slit_S4}
\end{figure*}

To examine the eruption of filament after merging, we have fixed the slit S$_4$ as shown in Figure~\ref{fig_slit_S4}.  
From the time-distance plot shown on the figure, it is evident that the eruption started at $\sim$ 05:10~UT, just after the F$_2$ and F$_3$ merging.
Further from the image, it is noticeable that the eruption had three phases. The first phase, starting at
$\sim$ 05:10 UT was a slow phase with a speed of $\sim$ 4~km~s$^{-1}$ and lasted up to 07:00 UT. After 07:00~UT the second phase started, where the eruption speed increased to about 17 km s$^{-1}$. The second phase lasted up to 09:00 UT. Another interesting feature observed in this phase was the observation of oscillations during the filament eruption. This oscillatory behaviour was observed throughout the second phase. Finally, the third phase started, where the eruption speed was maximum of $\sim$ 138 km s$^{-1}$. The eruption of this filament produced a large partial halo CME.

\medskip
\subsection{Magnetic field analysis} 

To examine the photospheric magnetic field evolution we performed a careful inspection of the HMI magnetic field movies (see online movies) and found regions of small-scale flux emergence/cancellation at the F$_1$ filament site and much more flux cancellation at the location of F$_2$ and F$_3$ filaments. 

\begin{figure*}   
\centering
\includegraphics[width=\textwidth]{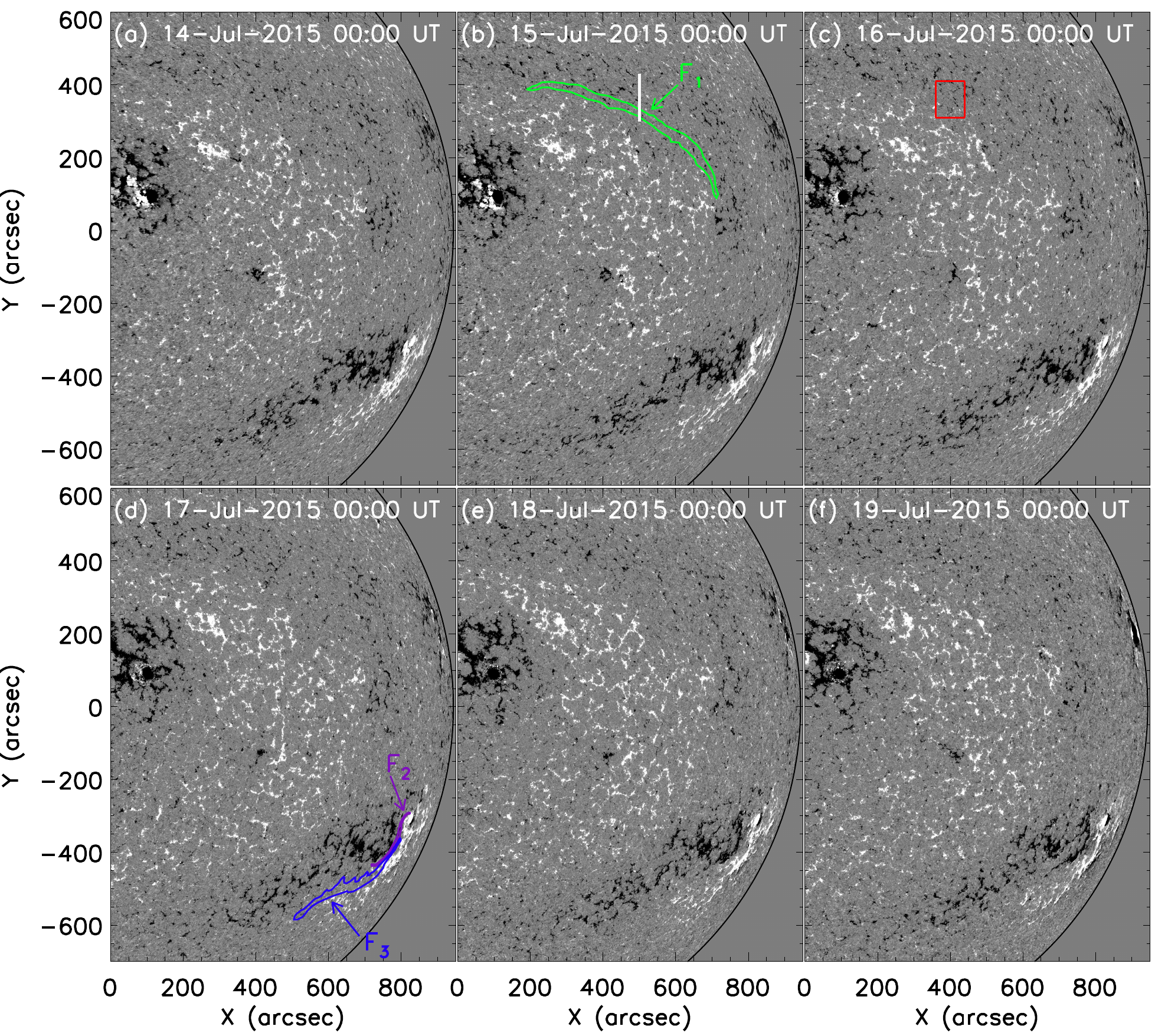}
\caption{Evolution of HMI LOS magnetic field during 14 -- 19 July 2015 before the onset of the eruption. The locations of the filaments are overlaid in `b' and `d' panel of the figure. The white line in panel `b' denotes the slit position, used for time-slice plot shown in Figure~\ref{ts_hmi}. The red box in panel `c' marks the region used for magnetic flux analysis, shown in Figure~\ref{mf_F1}. See also the accompanying movie.}
\label{hmi}
\end{figure*}

Figure~\ref{hmi} presents the magnetic field evolution between 14 -- 19 July 2015, with overploted the filament positions in panels `b' and `d'. The analyzed region,  which is located at $\sim 41 \degree$ from the central meridian, is marked by red box in panel `c' of the Figure~\ref{hmi}.

\begin{figure*}  
\centering
\includegraphics[width=\textwidth]{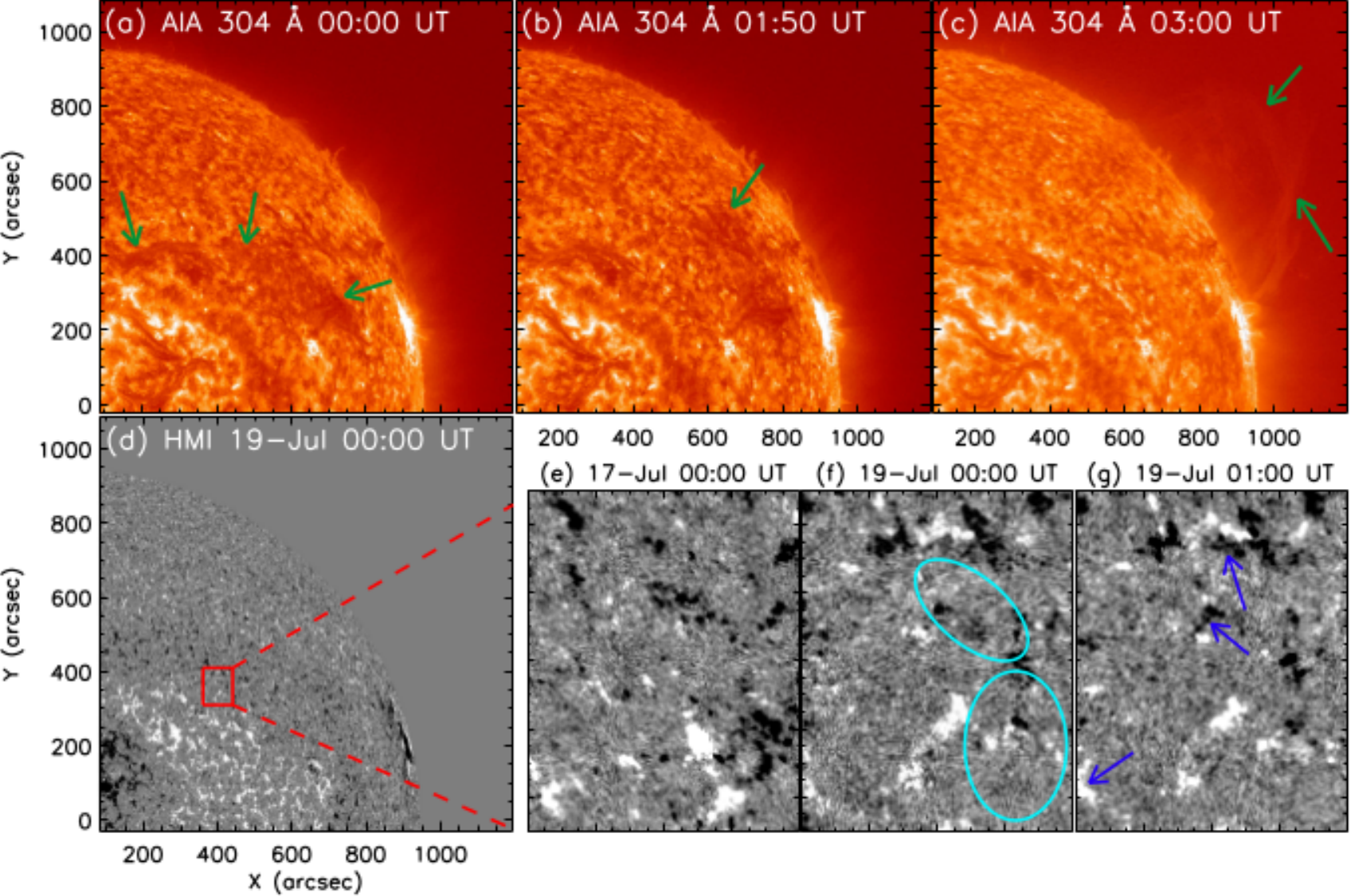}
\caption{(a)--(c) F$_1$ eruption evolution in AIA 304 \AA\ . The green arrows point to the rising F$_1$ filament. (d) The co-aligned HMI LOS magnetic field image. (e)--(g) The close-up magnetic field evolution, zoomed into the area shown as red rectangle in (d): Magnetic field quite state (e); During the magnetic field pre-eruptive state the cancellation is shown by cyan ellipses in panel (f) and flux emergence is shown by blue arrows in panel (g).}
\label{fig8}
\end{figure*}

The photospheric magnetic fields shown in Figure~\ref{hmi} indicate that F$_1$ is lying down along the neutral line.
The F$_1$ eastern part is ending in the bipolar region 
(500$\arcsec$, 371$\arcsec)$, where at 00:42~UT an emerging flux (EF) was observed. At the same time {\it i.e.} at 00:42 UT, the filament F$_1$ started to erupt and part of its material lifted up in northwest direction up to 03:20~UT, when it escaped the AIA FOV. Moreover, between 01:30~UT and 03:20~UT we observed a part of F$_1$ material to move down in southward direction up to the F$_2$ position.
In Figure~\ref{fig8} are shown a co-aligned image of AIA 304 (with a rising F$_1$) and LOS HMI magnetic field image. The close-up magnetic field evolution, zoomed into the area shown as red rectangle Figure~\ref{hmi} (c), is presented for the magnetic field quiet state Figure~\ref{fig8} (e), while during its pre-eruptive state (Figure~\ref{fig8} f, g), the cancellations and emergence of the magnetic field are shown.

\begin{figure*}              
\centering
\includegraphics[width=\textwidth]{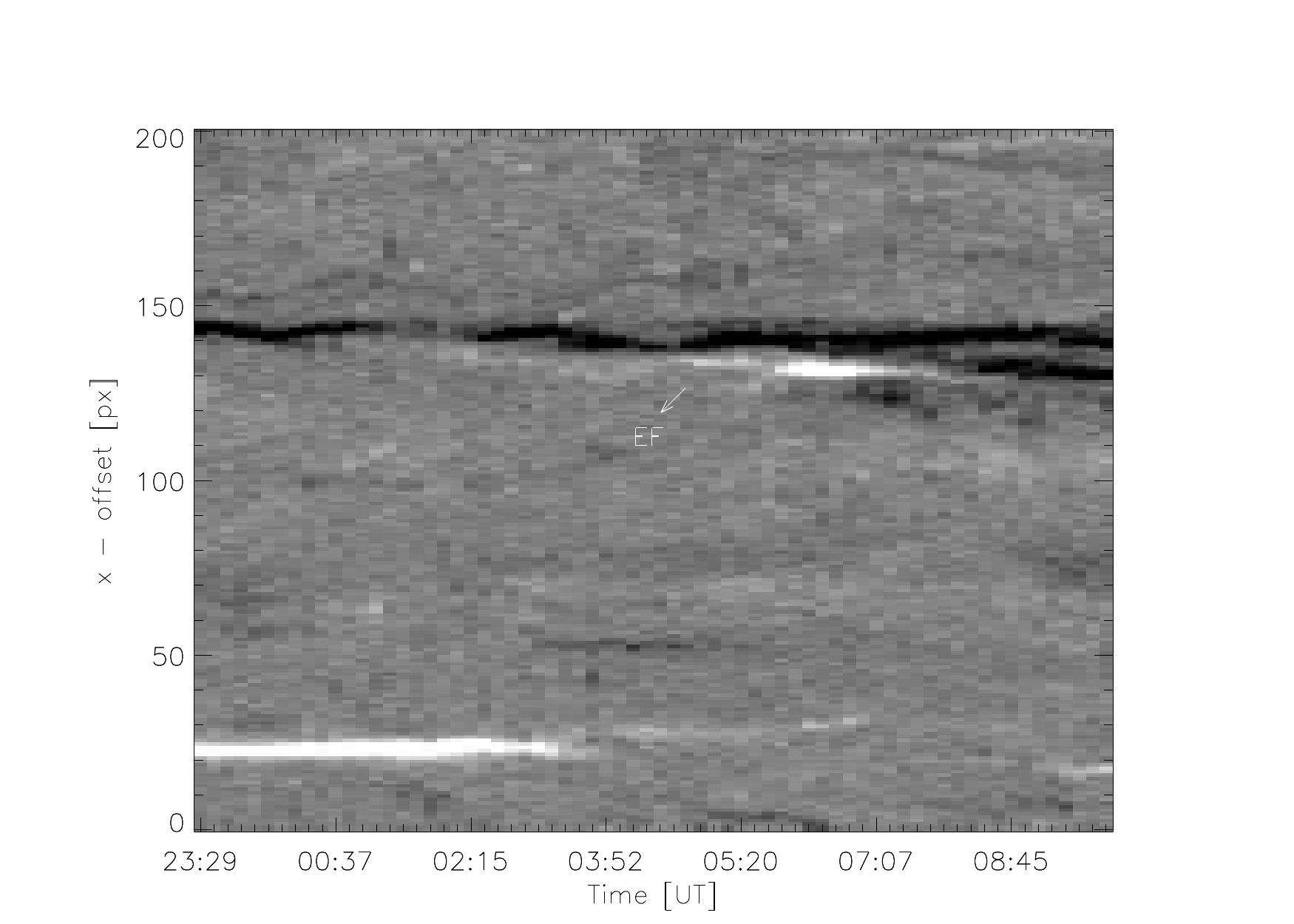}
\caption{Time-distance plot, showing the emerging flux close to the F$_1$ location. The slit position is shown in Figure~\ref{hmi}(b).}
\label{ts_hmi}
\end{figure*}

\begin{figure*}                    
\centering
\includegraphics[width=\textwidth]{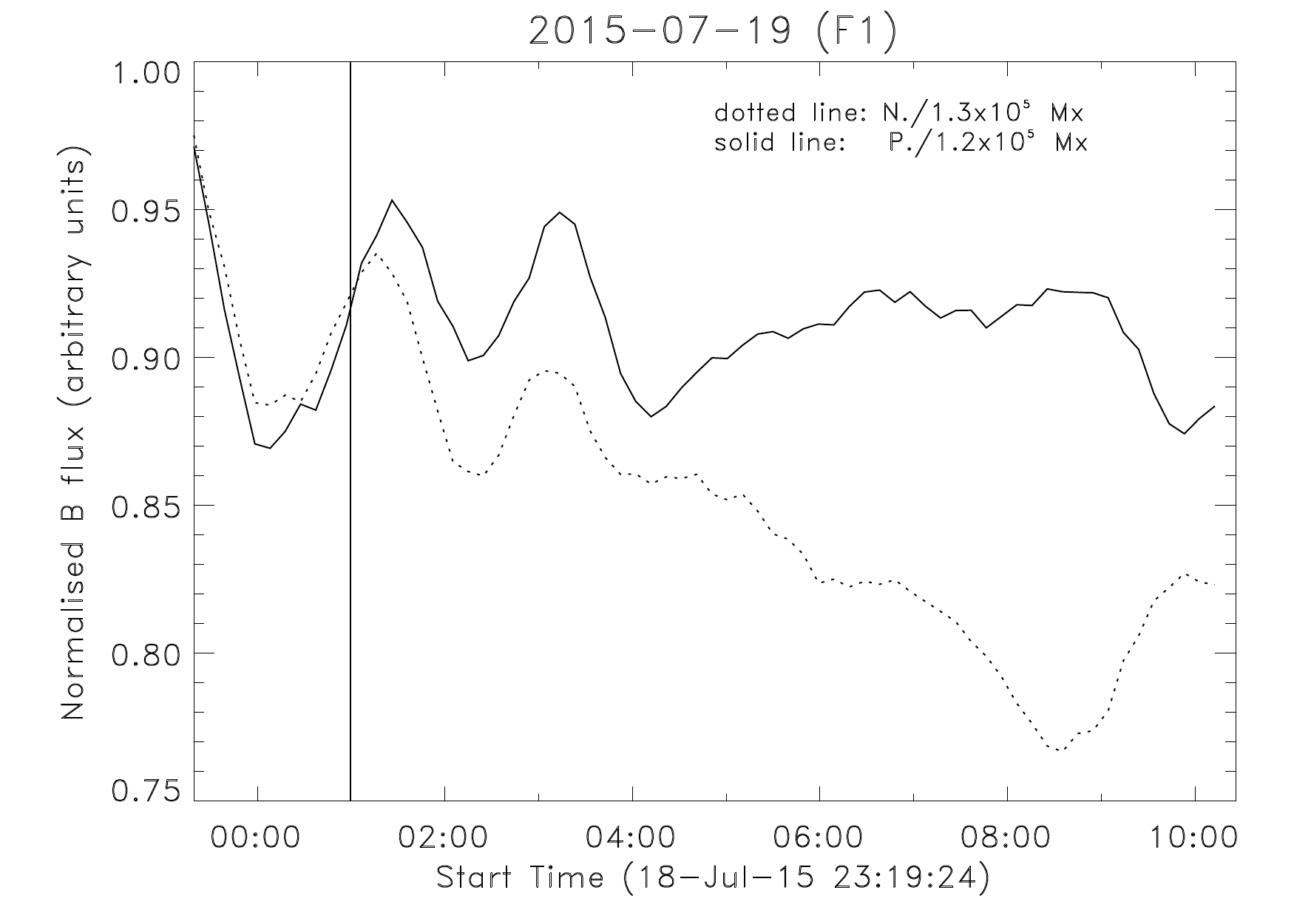}
\caption{Light curves of the normalised (to the maximum) magnetic fluxes, obtained from the red box, shown in Figure~\ref{hmi} c.  The vertical line represents the onset time of F$_1$ rising.}
\label{mf_F1}
\end{figure*}

The emerging flux close to the F$_1$ location is well visible in time-distance plot, shown in Figure~\ref{ts_hmi}.  The slit position is shown in Figure~\ref{hmi}(b) by white vertical line.
The time variations of both positive and negative LOS magnetic fluxes in the cancellation region, estimated in the box area shown in Figure~\ref{hmi}(c) from 23:19~UT on 18 July to 10:19~UT on 19 July, are presented in Figure~\ref{mf_F1}. During the pre-eruptive phase of the F$_1$ eruption, the positive flux steeply decreased up to 00:00~UT, when was the start time  of EF close the eastern F$_1$ footpoint (see Figure~\ref{fig8}) Then, it undergone a small increasing up to the F$_1$ eruption onset. After the eruption, the positive flux gradually decreased with an amplitude oscillations. During the pre-eruptive phase, the negative magnetic flux showed a similar behavior.  After the eruption it gradually decreased with an amplitude oscillations up to 04:30~UT and then the negative flux steeply decreased up to 09:00~UT. In the pre-eruption phase ({\it i.e.} 18 July 23:30 UT -- 19 July  00:00 UT), the decrease in both positive and negative fluxes can be considered as evidence of flux cancellation at the neutral line of the filament (\opencite {2010A&A...521A..49S,2011A&A...526A...2G}). Hence, there were sites of flux emergence and cancellation in and around the filament, influencing its stability. 

\begin{figure*}                 
\centering
\includegraphics[width=\textwidth]{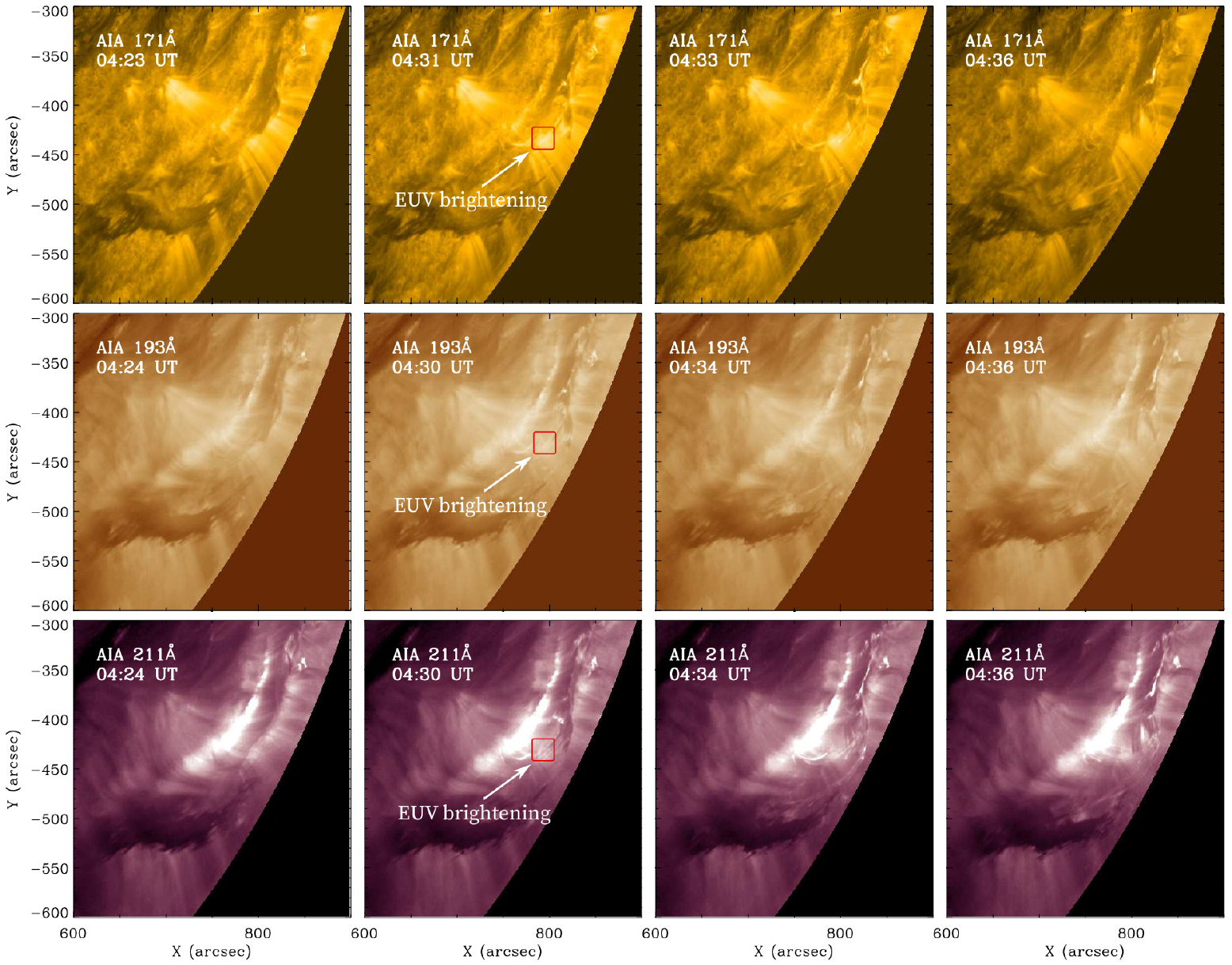}
\caption{Evolution of the EUV brightening in the vicinity of  F$_2$ and  F$_3$ filaments in three different AIA channels. Top panel: AIA 171~\AA; middle panel: AIA 193~\AA; bottom panel:  AIA 211~\AA. The 211 images are brightened with accent to the ribbon flare. All data are derotated up to 04:58 UT.}
\label{euv_br}
\end{figure*}

The pre-eruptive EUV brightening was observed in the vicinity of the filament channel prior to the  F$_1$ eruption, between 23:40~UT on July 18 and 00:50~UT on July 19 (see Figure \ref{fig_evolution304_171}). This pre-eruptive brightening represents series of small-scale patches aligned to the neutral line beneath the filament. The brightening was visible not only in 304 \AA\ and 171 \AA\ channels, but also in the high temperature AIA channels 193 \AA, 131 \AA\ and 94 \AA. The flux cancellations presented prior to the F$_1$ eruption are probably caused by the slow magnetic reconnections between the moving negative fluxes and its nearby positive fluxes (\opencite {1993SoPh..143..119W}), which would lead to some small-scale activities observed as the EUV brightenings, for example (\opencite {2019ApJ...871..229C}).

The pre-eruptive brightening in the vicinity of F$_2$ and F$_3$ was visible in all AIA channels. In Figure~\ref{euv_br} the evolution of EUV brightening is presented in three high temperature AIA channels, such as 171~\AA, 193~\AA\ and 211~\AA. 
We found that the first indications of the pre-eruptive brightening enhancement occurred after 02:40~UT. This brightening enhancement was slow and fragmented, i.e. in different small-scale locations of filament vicinity along the PIL. Such brightening could be caused by a slow reconnection acting in the course of the flux cancellations process during the pre-eruptive phase (\opencite {2019ApJ...871..229C}). After 04:00~UT significant and dynamic changes in the brightening occurred. The brightening enhancement covered the all filament vicinity, rapidly increasing and reaching a peak at 04:30~UT, i.e. when an extreme brightening was observed in some parts of the interacting and merging F$_2$ and F$_3$ flux ropes (FRs).
Afterward, the brightening rapidly decreased and after 05:00~UT, when the F$_2$--F$_3$ compound flux rope rose up, it returned to the initial rate.

\begin{figure*}                 
\centering
\includegraphics[width=\textwidth]{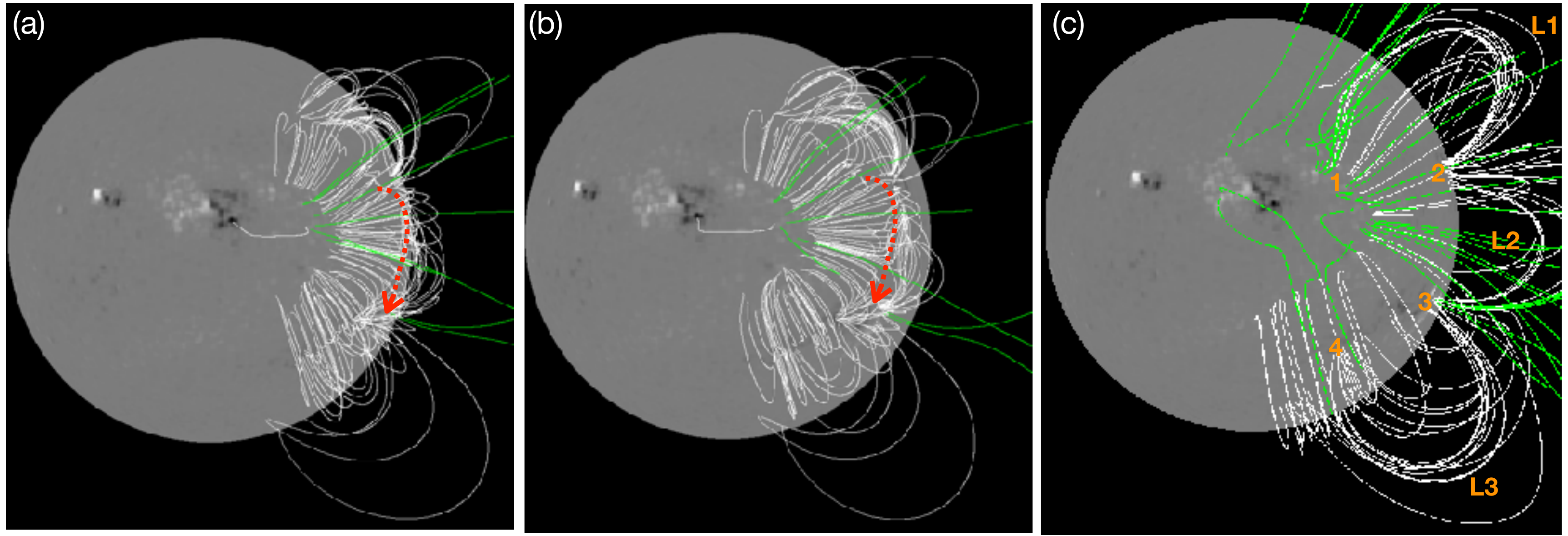}
\caption{PFSS extrapolation on 19 July 2015 at 00:00 UT (a). In panel (b) the PFSS extrapolation is shown from different view, rotated to the top of the moving channel.  The direction of ejected F$_1$ material towards F$_2$ and F$_3$ location is shown by red dashed arrow. The two filament locations are connected by different loop systems and represented by L$_1$, L$_2$, and L$_3$ in panel `c' of the figure.  }
\label{fig_pfss}
\end{figure*}

\section{Discussion and Summary}
\label{discussion}

We analyze the sympathetic eruption of three filaments observed on 19 July 2015. The filament F$_1$ was a quiet filament  located in the northern hemisphere, while the F$_2$ and F$_3$  filaments were located in the active region NOAA AR 12384 in the southern hemisphere.
The main results of this study are summarized as follows:

\begin{itemize}
\item {All the eruptions are sympathetic and  are associated with CMEs.} \\

\item{The time-distance analysis and the morphology of the filaments suggest that filament F$_1$ triggered F$_2$, which consequently triggered F$_3$}.\\

\item{We found that flux emergence/cancellation plays an important role in the observed filament eruption.  We suggest that the emergence/cancellation of magnetic fluxes near the  F$_1$ causes the flux rope to rise.} \\

\item{In addition to these processes, the material movement from F$_1$ to the F$_2$-F$_3$ location can additionally contribute to the F$_2$--F$_3$ compound FR destabilization.}\\

\item{ Our observations can be explained by the combination of  models proposed by \inlinecite{Ding2006} and \inlinecite{Torok2011}.}

\end{itemize}

Our analysis of the magnetic flux evolution beneath the eastern part of filament F$_1$ suggest that the flux emergence via magnetic flux cancellation caused F$_1$ destabilization. An important feature of the F$_1$, F$_2$ and F$_3$ evolution is the pre-eruptive EUV brightening. During the slow rise of filaments F$_2$ and F$_3$ an extreme EUV brightening occurred at some parts of the merging F$_2$ and F$_3$ (see  Figure~\ref{euv_br}), that suggests a sequence of partial merging episodes. Such brightening could be caused by the plasma heating, due to the energy released from reconnection site, below the rising prominence (\opencite {2015ApJ...807..144S}).

The merging of F$_2$ and F$_3$ is due to the stronger instability of the lower F$_2$ FR in comparison to those of the upper F$_3$ FR, which according to \opencite {2014ApJ...792..107K} is the condition for FRs merging mechanism to work. 
 Another condition for the merging of the two filaments could be the magnetic flux cancellation between them. Such a condition for the two filament merging was discussed in the study of \inlinecite{Chandra2011}. They found that the continuous decrease in the magnetic flux between the filaments brings them close to each other for the merging. Later the same observations were simulated by \inlinecite{Torok2011} who confirmed these observational results.

About the eruption of F$_2$--F$_3$ compound FR, it is important to note the suggestion of \opencite {2010ApJ...708..314A} that flux cancellation and tether-cutting reconnection are a key pre-eruption mechanisms for the buildup and the slow rise of an MFR, but they cannot trigger solar eruption alone. Moreover, the authors suggest the torus instability as an additional destabilizing mechanism.

Another event that can also cause significant EUV brightening and subsequently to affect the F$_2$ stability is the inflow of F$_1$ material in the vicinity of F$_2$. 
Such a brightening is considered as an observational signature of falling material and its impact on the solar atmosphere
(\opencite {2013ApJ...776L..12G}). Moreover, the fluid instabilities associated with the falling material were described recently by \opencite {2012A&A...540L..10I}. There are two mechanisms, compression and reconnection, that can explain the EUV brightening observed in the SDO/AIA channels. Which of them is responsible or dominant, depends on the amount of the energy associated with the observed emission. As \opencite {2013ApJ...776L..12G} pointed out, although the dominance of one mechanism over the other, both are likely occurring, since the falling material undoubtedly carries frozen-in magnetic flux. Therefore, the falling F$_1$ material in the vicinity of F$_2$ could be considered as additional mechanism that also facilitates the destabilization of F$_2$ filament. Moreover, this event provides an observational evidence for the physical linkage between the three eruptions.

\inlinecite{Ding2006} proposed the 2.5D MHD model for the sympathetic eruption.
They consider three FRs embedded in different arcade system
and same large scale magnetic field contributes for all these FRs systems. When one
FR becomes unstable (or erupted) due to catastrophic behaviour, the global shared
magnetic filed changes significantly for the unfinished FRs. As a result,
these undisturbed FRs becomes catastrophically unstable and erupt.

The 3D MHD simulation was proposed by \inlinecite{Torok2011} for the sympathetic
eruption of 1 August 2010. According to their model two mechanism are proposed.
The first FR erupted due to the converging flow and the another two FR are
initiated by the removal of magnetic flux above the FRs due to the magnetic
reconnection  triggered by first FR eruption.

In current observations, we believe that the first filament F$_1$ eruption could allow to reconnect the open field lines with the overlying field lines of F$_2$ and F$_3$ filament system.
As a result of this the magnetic tension above  F$_2$ and F$_3$ filament becomes weaker and they start to erupt. After comparing our reported observations with the above models, we believe our events can be  explained by the combination of both above discussed models. 

To explain the possible scenarios of the observations, we have performed the PFSS extrapolation of the pohospheric magnetic filed.
The result is presented in Figure \ref{fig_pfss}. 
In panel (a) the PFSS extrapolation on 19 July 2015 at 00:00~UT is presented. In panel (b) the PFSS extrapolation is shown from different view, rotated to the top of the moving channel. 
Using this figure, we present the following explanation for the current eruption:
The erupted part of the filament F$_1$ partially went through the open field lines (shown by green color in the figure) in north-west direction, which later observed as a CME. Major part of the erupted filament F$_1$ went towards the location of filaments F$_2$ through the channel of closed field lines (shown by red arrow). This part disturbed the filament F$_2$ and allows it to erupt. 
Since the erupted material probably was channeled under the closed magnetic fields, the observed case can be similar to the scenario of \inlinecite{2016ApJ...827L..12W}, {\it i.e.}, the filament (F$_1$) does not completely erupt under the closed field lines, but trigger the filament (F$_2$) nearby the open fields to erupt.
Another scenario of the sympathetic eruption can be explained by panel `c' of Figure \ref{fig_pfss}. In this scenario filament F$_1$ and F$_2$-F$_3$ are connected by following set of loops: Loop system L$_1$ connects  site `1' to site `2', Loop system L$_2$ connects site `2' to site `3' and similarly site `3' is connected to site `4' by loops L$_3$. 
The erupted material from F$_1$ went towards F$_2$-F$_3$ through loop system L$_1$, L$_2$, and L$_3$ and it disturbed the stability of filaments F$_2$ and F$_3$. As a result of this disturbance, filament F$_2$ and F$_3$ erupted.

\medskip

\noindent {\bf Disclosure of Potential Conflicts of Interest}  The authors declare that they have no conflicts of interest.
\begin{acks}
\noindent We are thankful to the referee for the constructive comments and suggestions. We acknowledge the open data policy of SDO, SOHO and GONG missions. This work is supported by the Bulgarian Science Fund and the Department of Science \& Technology, Government of India Fund under Indo-Bulgarian bilateral project No. DST/INT/BLS/P-11/2019 and KP-06-India/14 (19 -Dec-2019). PD thank the CSIR, New Delhi, for providing the research fellowship.

\end{acks}

\mbox{}~\\
\bibliographystyle{spr-mp-sola}
\bibliography{reference}
\IfFileExists{\jobname.bbl}{} {\typeout{}
\typeout{***************************************************************}
\typeout{***************************************************************}
\typeout{** Please run "bibtex \jobname" to obtain the bibliography}
\typeout{** and re-run "latex \jobname" twice to fix references}
\typeout{***************************************************************}
\typeout{***************************************************************}
\typeout{}}
\end{article} 
\end{document}